\documentclass[lettersize,journal]{IEEEtran}
\usepackage{amsmath,amsfonts}
\usepackage{algorithmic}
\usepackage{algorithm}
\usepackage{array}
\usepackage{textcomp}
\usepackage{stfloats}
\usepackage{url}
\usepackage{verbatim}
\usepackage{graphicx}
\usepackage{cite}
\usepackage{pifont}
\usepackage{multirow}
\usepackage[table,xcdraw]{xcolor}
\usepackage{colortbl}

\usepackage[caption=false,font=normalsize,labelfont=sf,textfont=sf]{subfig}
\usepackage[capitalize]{cleveref}
\crefname{section}{Section}{Sections}
\crefname{figure}{Figure}{Figures}
\crefname{table}{Table}{Tables}
\crefname{equation}{Equation}{Equations}

\begin{document}

\title{MarsSQE: Stereo Quality Enhancement for Martian Images Using Bi-level Cross-view Attention}

\author{Mai~Xu,~\IEEEmembership{Senior~Member,~IEEE,}
Yinglin Zhu,
Qunliang~Xing,~\IEEEmembership{Graduate~Student~Member,~IEEE,}
Jing~Yang,
and~Xin~Zou%
% \thanks{This work was supported by NSFC under Grants 62206011, 62250001, 62231002, and 62372024, and in part by the Beijing Natural Science Foundation under Grant L223021. \textit{(Corresponding author: Mai Xu.)}}
\thanks{Mai Xu, Yinglin Zhu, Qunliang Xing, and Jing Yang are with the School of Electronic Information Engineering, Beihang University, Beijing 100191, China. Yinglin Zhu and Qunliang Xing are also with the Shen Yuan Honors College, Beihang University, Beijing 100191, China (e-mail: maixu@buaa.edu.cn; zhuyinglin@buaa.edu.cn; ryanxingql@gmail.com; jing\_yang@buaa.edu.cn). \textit{(Corresponding author: Qunliang Xing.)}}%
\thanks{Xin Zou is with the Beijing Institute of Spacecraft System Engineering, Beijing 100094, China (e-mail: zouxin501@163.com).}%
}

\markboth{MarsSQE: Stereo Quality Enhancement for Martian Images Using Bi-level Cross-view Attention}%
{Zhu~\MakeLowercase{\textit{et~al.}}: MarsSQE: Stereo Quality Enhancement for Martian Images Using Bi-level Cross-view Attention}


\maketitle

\begin{abstract}
    Stereo images captured by Mars rovers are transmitted after lossy compression due to the limited  bandwidth between Mars and Earth. Unfortunately, this process results in undesirable compression artifacts. In this paper, we present a novel stereo quality enhancement approach for Martian images, named MarsSQE. First, we establish the first dataset of stereo Martian images. Through extensive analysis of this dataset, we observe that cross-view correlations in Martian images are notably high. Leveraging this insight, we design a bi-level cross-view attention-based quality enhancement network that fully exploits these inherent cross-view correlations. Specifically, our network integrates pixel-level attention for precise matching and patch-level attention for broader contextual information. Experimental results demonstrate the effectiveness of our MarsSQE approach.
\end{abstract}


\section{Introduction}

\IEEEPARstart{D}{riven} by the rapid advancements in Mars exploration, rovers such as Perseverance~\cite{makimars2020} and Zhu Rong~\cite{liangnavigation2021} have successfully landed on Mars and captured invaluable images showcasing the Martian surface.
These images provide invaluable data for scientific research.
However, the vast communication distance between Earth and Mars---reaching up to 400 million kilometers~\cite{manzeyhuman2004}---poses a major challenge for transmitting these images.
To overcome this challenge, Martian images are typically subjected to lossy compression~\cite{hayespreflight2021}, which inevitably introduces compression artifacts and degrades image quality.
This highlights the need for Martian image quality enhancement.

To the best of our knowledge, there is only one pioneering study on the quality enhancement of Martian images~\cite{liumarsqe2024}, which leverages semantic similarities among Martian images.
In addition, approaches for enhancing Earth image quality offer straightforward solutions for improving Martian image quality.
For instance, Dong~et~al.~\cite{dongcompression2015} introduced the first Convolutional Neural Network (CNN) for quality enhancement, proposing a four-layer network named the Artifacts Reduction CNN (AR-CNN).
Zhang~et~al.~\cite{zhanggaussian2017} developed a Denoising CNN (DnCNN) capable of removing blocking effects caused by JPEG compression.
Further advancements have focused on blind and resource-efficient quality enhancement~\cite{fujpeg2019,guoconvolutional2019,xingearly2020,lilearning2020,xingdaqe2023}.

The aforementioned quality enhancement approaches are all monocular-based, relying solely on single-view images.
However, Mars rovers are equipped with stereo cameras to capture binocular images for depth estimation and navigation~\cite{belliiimars2017,bellmars2021}.
These stereo images exhibit cross-view correlations unavailable in single-view images, making them valuable for quality enhancement tasks.
In fact, studies on Earth image quality enhancement have demonstrated the potential of cross-view information exchange.
For instance, PASSRnet~\cite{wanglearning2019} and iPASSR~\cite{wangsymmetric2021} utilize cross-view information for stereo super-resolution, achieving superior results compared to monocular approaches.

Despite this potential, stereo quality enhancement for Martian images remains unexplored.
To address this gap, we establish the first stereo Martian image dataset for quality enhancement.
Through extensive analysis, we confirm that these images exhibit notably high intra-view and cross-view correlations.
Motivated by these insights, we propose a novel stereo quality enhancement approach for Martian images, named MarsSQE.
Our approach incorporates a bi-level cross-view attention-based network that fully exploits these inherent correlations.
In addition to employing pixel-level attention, commonly used for Earth images~\cite{wanglearning2019,wangsymmetric2021}, our network integrates patch-level attention to capture broader contextual information.
This bi-level design is particularly important because the Martian surface is highly unstructured, featuring irregular gravel, rocks, soil, and dunes~\cite{liuhybrid2023}, which challenges precise pixel-level matching~\cite{sunloftr2021}.
Moreover, accurate cross-view correspondence depends not only on individual pixels but also on their spatial relationships with surrounding pixels, as verified by~\cite{sunloftr2021}.

Finally, we conduct extensive experiments and validate the effectiveness of our MarsSQE approach. 
The contributions of this paper are summarized as following:
\begin{itemize}
    \item {We establish and analyze the first dataset of stereo Martian images, highlighting their high inherent intra-view and cross-view correlations.}
    \item {We propose the first stereo quality enhancement approach for Martian images, integrating bi-level cross-view attention to effectively exploit these correlations.}
\end{itemize}

\section{Stereo Martian Image Dataset}

\subsubsection*{\textbf{Dataset establishment}}

\begin{figure*}[t]
    \centering
    \includegraphics[width=\textwidth]{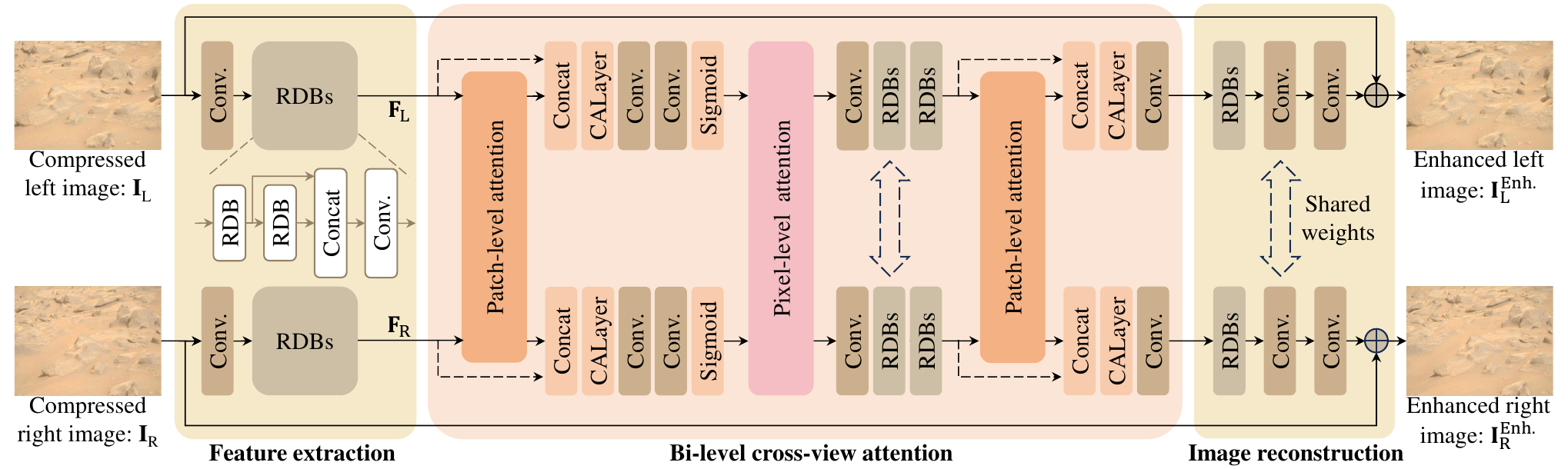}
    \caption{Overall framework of the proposed MarsSQE approach.}
    \label{model1}
\end{figure*}

\begin{figure}[t]
    \centering
    \includegraphics[width=\linewidth]{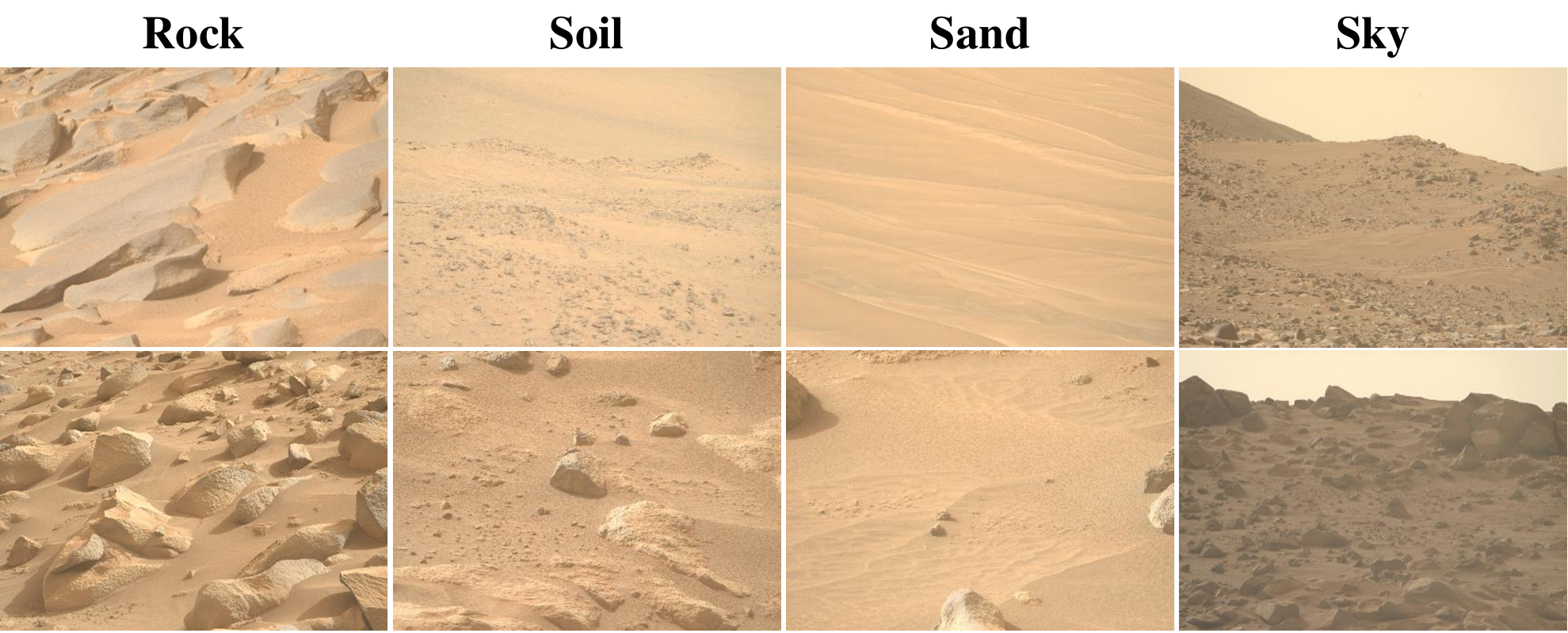}
    \caption{Sample images in our dataset of four different landforms.}
    \label{dataset}
\end{figure}

We establish the first stereo Martian image dataset for quality enhancement.
Our dataset comprises 1,350 pairs of stereo Martian images captured by the Mast Camera Zoom (Mastcam-Z)---an imaging system consisting of a pair of RGB cameras mounted on the Perseverance Rover~\cite{bellmars2021}.
These image pairs are all binocular with a resolution of \(1152 \times 1600\).
The dataset covers four primary Martian landforms: rock, soil, sand, and sky, as shown in \cref{dataset}.
When establishing the dataset, we exclude images with severe occlusion, corruption, or insignificant parallax shifts.
The images are of high quality without noticeable artifacts, serving as ground truth for our dataset\footnote{High-quality or even lossless Martian images are returned when downlink data volumes are high~\cite{bellmars2021} and thus are available to us.}.
Following the Mastcam-Z procedure, these high-quality images are then compressed using JPEG to emulate the actual image transmission process, resulting in compressed stereo Martian images.

\subsubsection*{\textbf{Intra-view and cross-view correlations of stereo Martian images}}

We compare Martian images from our dataset with Earth images from the Flickr1024 dataset~\cite{wangFlickr10242019}.
Then we calculate the Correlation Coefficient (CC) and Mutual Information (MI), with higher values indicating greater similarity.
Specifically, we pair patches within the same image to assess intra-view similarity, and between the left and right images to assess cross-view similarity.
As shown in \cref{compare}, Martian images exhibit significantly higher CC results than Earth images, suggesting stronger correlations both within and across views.
Additionally, Martian images show 21.66\% higher MI result for intra-view and 20.61\% higher for cross-view, indicating greater information gain in both contexts.
These results highlight the significance of the stereo quality enhancement paradigm for Martian images, as it leverages complementary stereo information to improve image quality.

\begin{table}[t]
    \caption{Intra-view and cross-view similarities of Martian and Earth stereo images}
    \label{compare}
    \centering
    \begin{tabular}{>{\centering\arraybackslash}p{1.4cm}|>{\centering\arraybackslash}p{0.5cm}|>{\centering\arraybackslash}p{1.5cm}|>{\centering\arraybackslash}p{1.5cm}}
    \hline
    \rowcolor{gray!20}
    \multicolumn{2}{c|}{Metric} & Martian & Earth\\
    \hline
    \hline
    \multirow{2}{*}{Intra-view} & CC & \bf{0.7719} & 0.0688\\
    & MI & \bf{1.1657} & 0.9582\\
    \hline
    \multirow{2}{*}{Cross-view} & CC & \bf{0.7738} & 0.0793\\
    & MI & \bf{1.1657} & 0.9665\\
    \hline
    \end{tabular}
\end{table}

\section{MarsSQE Approach}

\subsubsection*{\textbf{Overview}}

The overall framework of our MarsSQE approach is shown in \cref{model1}.
Our framework includes three procedures: feature extraction, bi-level cross-view attention, and image reconstruction.
Given the compressed image pair \({\mathbf{I}}_{\text{L}}\) and \({\mathbf{I}}_{\text{R}}\) as inputs, we first extract image features and send them to the proposed bi-level cross-view attention sub-network for enhancement.
This sub-network employs two patch-level attention modules and one pixel-level attention module, such that details missed in the current view but reserved in the other view during compression can be discovered.
Finally, we reconstruct images \({\mathbf{I}}_{\text{L}}^{\text{Enh.}}\) and \({\mathbf{I}}_{\text{R}}^{\text{Enh.}}\) from the enhanced features.
In the subsequent paragraph, we describe the pipeline of the bi-level cross-view attention sub-network.
Then, we detail the design of patch-level and pixel-level attention modules in this sub-network.

\subsubsection*{\textbf{Bi-level cross-view attention}}

Due to the unique unstructured terrain of Mars, Martian images are highly similar, which causes difficulties in stereo image matching.
Therefore, we first apply patch-level attention between two views with broader contextual information.
As shown in the middle of \cref{model1}, the resulting feature is concatenated with the input feature and fused by a Channel Attention (CA) layer~\cite{zhangimage2018} and a convolution layer with a kernel size of \(1 \times 1\).
The fusion of features is then passed through a convolution layer with a kernel size of \(3 \times 3\), followed by a sigmoid layer.
After that, a pixel-level attention module is used for precise matching between two views, followed by a convolution layer with a kernel size of \(1 \times 1\) and two Residual Dense Blocks (RDBs)~\cite{zhangresidual2018}.
Finally, the feature is enhanced by another patch-level attention module.

\begin{figure*}[t]
    \centering
    \includegraphics[width=\textwidth]{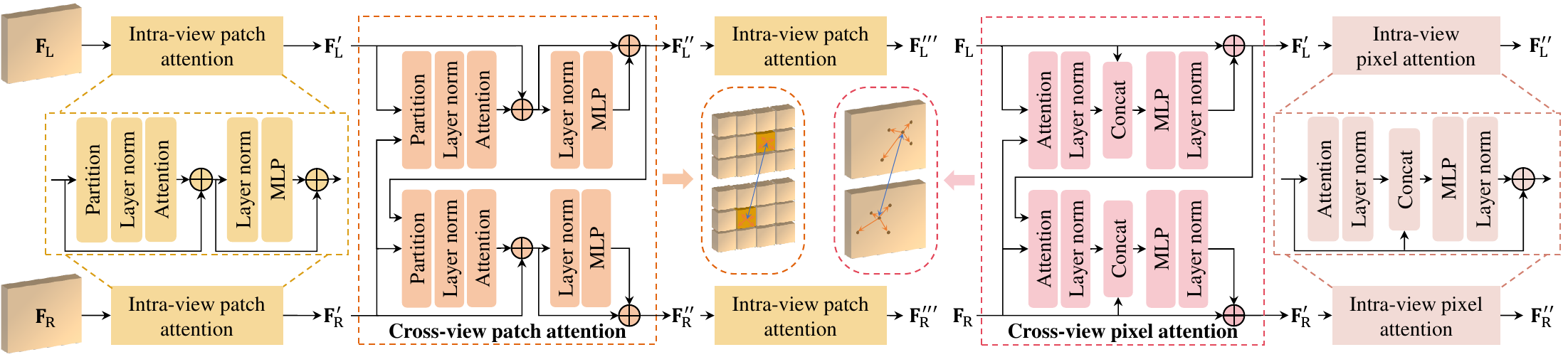}
    \caption{Illustration of the patch-level attention module (left) and the pixel-level attention module (right).}
    \label{model2}
\end{figure*}

\subsubsection*{\textbf{Patch-level attention module}}

As shown in \cref{model2} left, the patch-level attention module has two main operations: intra-view patch attention \( A_{\text{Patch}}^{\text{Intra}} \) and cross-view patch attention \( A_{\text{Patch}}^{\text{Cross}} \).
Mathematically, given input features $\mathbf{F}_{\text{L}}$ and $\mathbf{F}_{\text{R}}$, we obtain the enhanced features \( \mathbf{F}^{'''}_{\text{L}} \) and \( \mathbf{F}^{'''}_{\text{R}} \) by the following processes:
\begin{align}
    {\mathbf{F}}^{'}_{V} &= A_{\text{Patch}}^{\text{Intra}} \left( \mathbf{F}_{V} \right), V \in \{ \text{L}, \text{R} \},\\
    \mathbf{F}^{''}_{\text{L}}, \mathbf{F}^{''}_{\text{R}} &= A_{\text{Patch}}^{\text{Cross}} \left( \mathbf{F}^{'}_{\text{L}}, \mathbf{F}^{'}_{\text{R}} \right),\\
    {\mathbf{F}}^{'''}_{V} &= A_{\text{Patch}}^{\text{Intra}} \left( \mathbf{F}^{''}_{V} \right), V \in \{ \text{L}, \text{R} \}.
\end{align}
Take the left-view enhancement as an example.
Supposing $\mathbf{F}_{\text{L}}\in \mathbb{R}^{B\times C\times H\times W}$, the input feature is divided into non-overlapping patches of size $P_{h} \times P_{w}$.
In $A_{\text{Patch}}^{\text{Intra}}$, self-attention is performed in every patch, generating $\mathbf{Q}, \mathbf{K}, \mathbf{V} \in \mathbb{R}^{ \left( B \times \frac{H}{P_{h}} \times \frac{W}{P_{w}} \right) \times \left( P_{h} \times P_{w} \right) \times C}$ for each view.
Within the cross-view attention calculation, the left-view $\mathbf{Q}_{\text{L}}$ and the right-view $\mathbf{K}_{\text{R}}, \mathbf{V}_{\text{R}}$ take the following calculation:
\begin{equation}
    \text{Attention} \left( \mathbf{Q}_{\text{L}}, \mathbf{K}_{\text{R}}, \mathbf{V}_{\text{R}} \right) = \text{Softmax} \left( \mathbf{Q}_{\text{L}} \mathbf{K}_{\text{R}}^T \right) \mathbf{V}_{\text{R}}.
\end{equation}

\subsubsection*{\textbf{Pixel-level attention module}}

As shown in \cref{model2} right, we first employ cross-view pixel attention to the left-view feature \( \mathbf{F}^{'}_{\text{L}} \) and the right-view feature \( \mathbf{F}^{'}_{\text{R}} \), and then employ intra-view pixel attention to each view.
This way, we obtain the enhanced features \( \mathbf{F}^{''}_{\text{L}} \) and \( \mathbf{F}^{''}_{\text{R}} \).
The whole process can be formulated as:
\begin{align}
    \mathbf{F}^{'}_{\text{L}}, \mathbf{F}^{'}_{\text{R}} &= A_{\text{Pixel}}^{\text{Cross}} \left( \mathbf{F}_{\text{L}}, \mathbf{F}_{\text{R}} \right),\\
    \mathbf{F}^{''}_{V} &= A_{\text{Pixel}}^{\text{Intra}} \left( \mathbf{F}^{'}_{V} \right), V \in \{ \text{L}, \text{R} \},
\end{align}
where $A_{\text{Pixel}}^{\text{Cross}}$ and $A_{\text{Pixel}}^{\text{Intra}}$ represent cross-view pixel attention and intra-view pixel attention, respectively.

\section{Experiments}

\subsection{Experimental Settings}

We randomly select 800 image pairs from our stereo Martian image dataset for training, and 100 image pairs for testing.
Following the Mastcam-Z procedure, JPEG is utilized for Martian image compression, with the Quality Factor (QF) set to 30, 40, 50, and 60 separately.
We choose a patch size of 16 in the patch-level attention module.
The batch size is set to 4.
We use the Adam optimizer~\cite{kingmaadam2015} with \( \beta_{1}=0.9 \) and \( \beta_{2}=0.999 \).
The learning rate is set to \( 1\times10^{-3} \) initially and decreases by 0.9 times every three epochs.
Our network is trained with a maximum of two NVIDIA 4090 GPUs using the PyTorch framework.
The L1 loss is used for network training, which is formulated as:
\begin{equation}
    \mathcal{L}=\left \| \mathbf{I}_{\text{L}}^{\text{Enh.}}-\mathbf{I}_{\text{L}}^{\text{Raw}} \right \|_1+\left \| \mathbf{I}_{\text{R}}^{\text{Enh.}}-\mathbf{I}_{\text{R}}^{\text{Raw}} \right \|_1,
\end{equation}
where $\mathbf{I}_{\text{L}}^{\text{Enh.}}$ and $\mathbf{I}_{\text{R}}^{\text{Enh.}}$ represent enhanced left and right view images generated by our MarsSQE approach, and $\mathbf{I}_{\text{L}}^{\text{Raw}}$ and $\mathbf{I}_{\text{R}}^{\text{Raw}}$ represent their ground-truth raw images.

We compare our MarsSQE approach with several quality enhancement baselines, including AR-CNN~\cite{dongcompression2015}, DnCNN~\cite{zhanggaussian2017}, CBDNet~\cite{guoconvolutional2019}, RBQE~\cite{xingearly2020}, and MarsQE~\cite{liumarsqe2024}.
We also compare it to super-resolution baselines PASSRnet~\cite{wanglearning2019} and iPASSR~\cite{wangsymmetric2021}.
MarsQE is the only approach specifically designed for Martian images, while other approaches were originally proposed for Earth images.
Among these approaches, PASSRnet and iPASSR handle binocular images, while others are applicable to monocular ones.
All monocular approaches are retrained using 800 left-view and 800 right-view Martian images.
PASSRnet and iPASSR are also retrained with the scale factor set to 1.
Then we calculate the Peak Signal to Noise Ratio (PSNR) and Structural SIMilarity (SSIM) index for enhancement quality.

\subsection{Evaluation}

\begin{figure}[t]
    \centering
    \includegraphics[width=.83\linewidth]{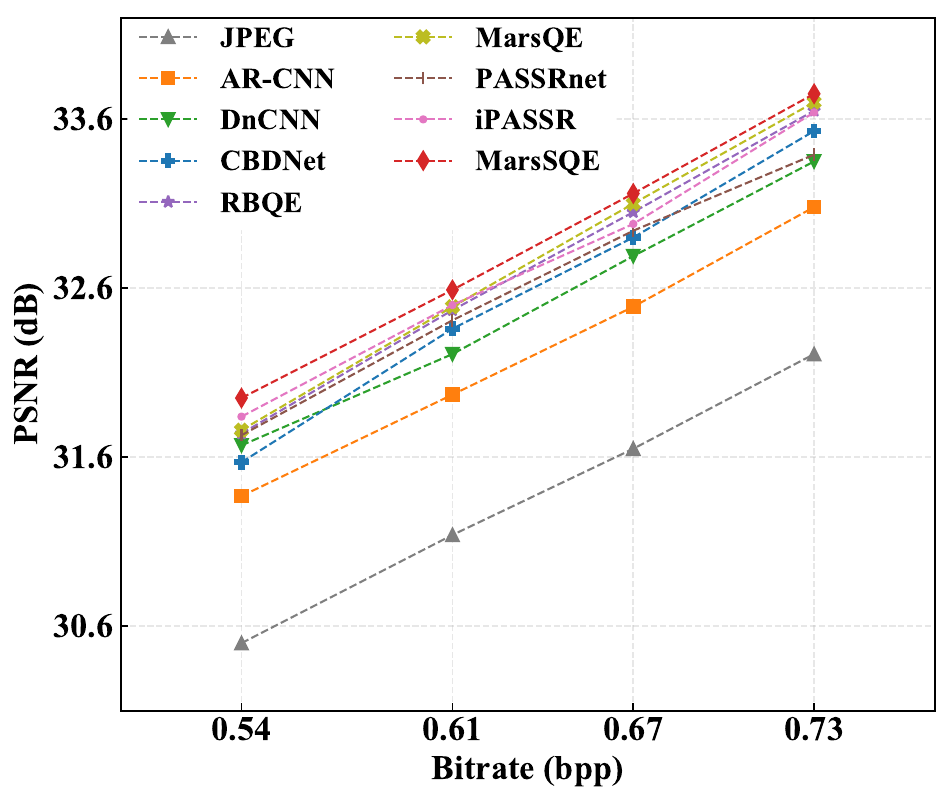}
    \caption{Rate-Distortion (RD) curves of MarsSQE and compared approaches. The rate is measured in bits per pixel (bpp).}
    \label{RD}
\end{figure}

\begin{table*}[t]
    \caption{Quantitative Comparison with PSNR/SSIM on Stereo Martian Image Dataset under QF=30, 40, 50, and 60. Best Results are in Bold}
    \label{psnr}
    \centering
    \begin{tabular}{c|c|c|c|c|c|c|c|c|c|c}
        \hline
        \rowcolor{gray!20}
        View & QF &JPEG& AR-CNN& DnCNN& CBDNet& RBQE& MarsQE& PASSRnet& iPASSR& MarsSQE\\
        \hline
        \hline
        \multirow{4}{*}{Left}
        & 30 & 30.44/0.823& 31.30/0.838& 31.60/0.845& 31.50/0.843& 31.66/0.847& 31.68/0.849& 31.64/0.847& 31.76/0.850& \textbf{31.86/0.852}\\
        & 40 & 31.06/0.843& 31.89/0.856& 32.14/0.862& 32.29/0.866& 32.38/0.868& 32.40/0.869& 32.32/0.867& 32.41/0.868& \textbf{32.50/0.870}\\
        & 50 & 31.57/0.858& 32.41/0.871& 32.71/0.877& 32.82/0.880& 32.95/0.882& 33.01/0.884& 32.84/0.880& 32.89/0.881& \textbf{33.07/0.885}\\
        & 60 & 32.14/0.873& 33.00/0.885& 33.27/0.891& 33.46/0.894& 33.55/0.896& 33.61/0.897& 33.30/0.892& 33.55/0.896& \textbf{33.66/0.898}\\
        \hline
        \hline
        \multirow{4}{*}{Right} 
        & 30 & 30.57/0.828& 31.44/0.844& 31.74/0.851& 31.63/0.849& 31.83/0.853& 31.84/0.854& 31.81/0.853& 31.93/0.855& \textbf{32.04/0.858}\\
        & 40 & 31.21/0.847& 32.05/0.862& 32.29/0.867& 32.43/0.871& 32.56/0.873& 32.58/0.875& 32.50/0.872& 32.59/0.874& \textbf{32.69/0.876}\\
        & 50 & 31.73/0.862& 32.57/0.876& 32.87/0.882& 32.97/0.884& 33.15/0.887& 33.20/0.889& 33.03/0.885& 33.06/0.886& \textbf{33.26/0.890}\\
        & 60 & 32.29/0.876& 33.15/0.889& 33.43/0.895& 33.61/0.898& 33.74/0.900& 33.79/0.901& 33.48/0.896& 33.74/0.900& \textbf{33.84/0.902}\\
        \hline
        \hline
        \multirow{4}{*}{Avg.} 
        & 30 & 30.50/0.825 & 31.37/0.841& 31.67/0.848& 31.57/0.846& 31.74/0.850& 31.76/0.851& 31.73/0.850& 31.84/0.852& \textbf{31.95/0.855}\\
        & 40 & 31.14/0.845 & 31.97/0.859& 32.21/0.865& 32.36/0.869& 32.47/0.870& 32.49/0.872& 32.41/0.869& 32.50/0.871& \textbf{32.59/0.873}\\
        & 50 & 31.65/0.860 & 32.49/0.873& 32.79/0.879& 32.90/0.882& 33.05/0.885& 33.10/0.886& 32.94/0.883& 32.98/0.883& \textbf{33.16/0.887}\\
        & 60 & 32.21/0.875 & 33.08/0.887& 33.35/0.893& 33.53/0.896& 33.65/0.898& 33.70/0.899& 33.39/0.894& 33.64/0.898& \textbf{33.75/0.900}\\
        \hline
    \end{tabular}
\end{table*}

\begin{figure*}[ht]
    \centering
    \subfloat[]{
        \includegraphics[width=.95\textwidth]{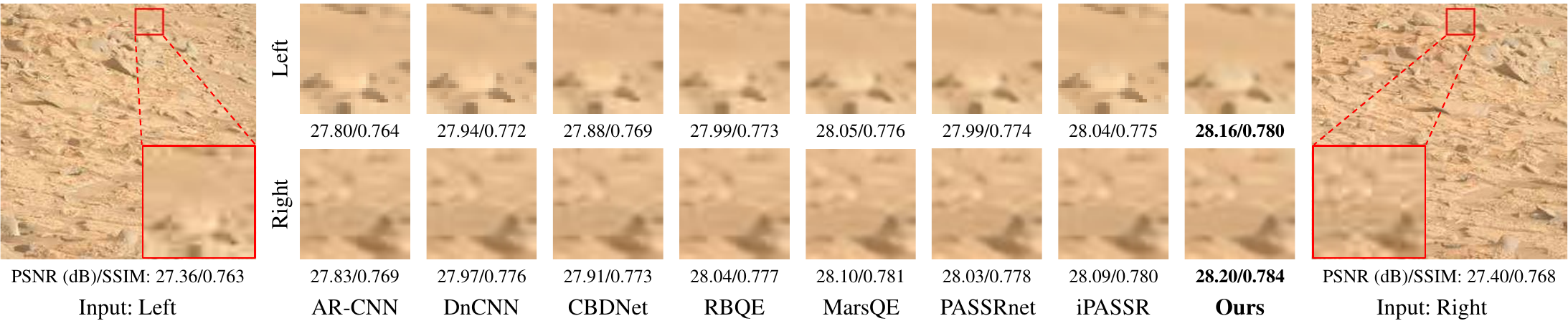}
        \label{qualitative:a}
    }\\
    % \vspace{-.7em}
    \subfloat[]{
        \includegraphics[width=.95\textwidth]{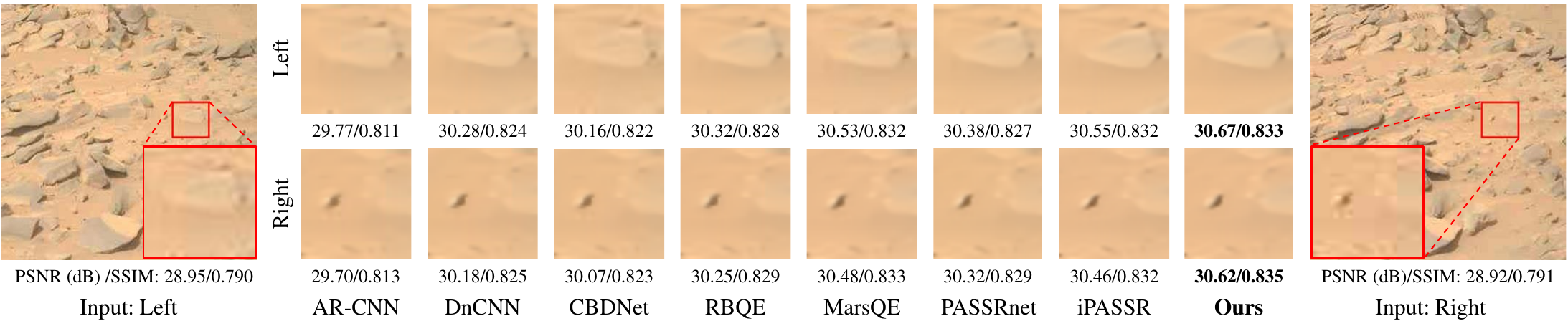}
        \label{qualitative:b}
    }
    \caption{Visual comparison of our MarsSQE and other approaches under QF=30.}
    \label{qualitative}
\end{figure*}

\begin{table}[t]
    \caption{Parameter number and PSNR (dB) of our MarsSQE and compared binocular approaches under QF=30}
    \label{Params}
    \centering
    \begin{tabular}{>{\centering\arraybackslash}p{1.5cm}|>{\centering\arraybackslash}p{1.4cm}|>{\centering\arraybackslash}p{1.4cm}|>{\centering\arraybackslash}p{1.5cm}}
    \hline
    \rowcolor{gray!20}
    Approach & PSNR& SSIM& Params.\\
    \hline
    \hline
    PASSRnet & 31.73& 0.8502& 1.36M\\
    iPASSR& 31.84& 0.8524& 1.37M\\
    \hline
    MarsSQE-S & 31.90& 0.8537& 1.00M\\
    MarsSQE-M & 31.92& 0.8539& 1.32M\\
    MarsSQE-L & 31.95& 0.8550& 1.69M\\
    \hline
    \end{tabular}
\end{table}

\subsubsection*{\textbf{Quantitative performance}}
As shown in \cref{psnr}, our MarsSQE approach achieves superior performance compared to other approaches.
Specifically, MarsSQE achieves a PSNR of 31.95 dB under QF=30, which is 0.19 dB higher than the best monocular approach, MarsQE, and 0.11 dB higher than the best binocular approach, iPASSR.
Similar trends are observed under QF=40, 50, and 60.
Furthermore, MarsSQE achieves the best Rate-Distortion (RD) performance, as shown in \cref{RD}.
In conclusion, MarsSQE outperforms both monocular and binocular approaches in quantitative evaluation.

\subsubsection*{\textbf{Qualitative performance}}
As shown in \cref{qualitative:a,qualitative:b}, the texture and boundary of rocks are blurred and noisy due to compression in both the left- and right-view images.
MarsSQE successfully restores the texture and distinguishes the boundary, while other approaches (such as PASSRnet in \cref{qualitative:a} and CBDNet in \cref{qualitative:b}) fail to recover the complete rock.
Additionally, some approaches introduce artifacts, such as AR-CNN and DnCNN in both images.
In conclusion, MarsSQE provides the best qualitative enhancement performance.

\subsubsection*{\textbf{Efficiency performance}}
We provide two lightweight variants, MarsSQE-M and MarsSQE-S, by reducing the number of main channels in MarsSQE(-L) from 64 to 48 and 32, respectively.
As shown in \cref{Params}, MarsSQE-S surpasses iPASSR by 0.06 dB while requiring 27\% fewer parameters.
MarsSQE-M, with fewer parameters than both iPASSR and PASSRnet, provides a PSNR gain of 0.08 dB.
Additionally, MarsSQE-L, despite having 23\% more parameters than iPASSR, delivers a 0.11 dB higher PSNR.
These results demonstrate the superior efficiency of MarsSQE compared to existing approaches.

\subsection{Ablation Study}

To verify the effectiveness of the core component of our MarsSQE approach, i.e., the bi-level cross-view attention, we perform an ablation study.
Specifically, both levels of cross-attention are removed, and only several convolutional layers and residual dense blocks are maintained to extract and reconstruct compressed images.
The results show that PSNR is reduced by 0.18 dB (but still higher than all monocular approaches), which proves that cross-view information is very effective for binocular image recovery.
This result further demonstrates the importance of conducting stereo quality enhancement for Martian images.

\section{Conclusion}

In this letter, we established the first stereo Martian image dataset.
By evaluating the correlations between left and right views, we found that cross-view relationships are significantly stronger in Martian images compared to Earth images.
This motivated us to propose a bi-level cross-view attention-based stereo quality enhancement network for Martian images.
Our network integrates patch-level attention for broader contextual information and pixel-level attention for precise matching to fully exploit the inherent cross-view correlations.
Experiments demonstrate that our MarsSQE approach achieves better performance in both quantitative and qualitative comparisons.

\bibliographystyle{IEEEtran}
\bibliography{references}
\vfill

\end{document}